# Database Optimization to Recommend Software Developers using Canonical Order Tree

T.M. Amir-Ul-Haque Bhuiyan*, Mehedi Hasan Talukdar*, Ziaur Rahman[†], Dr. Mohammad Motiur Rahman*
*Department of Computer Science and Engineering, [†]Department of Information and Communication Technology
Mawlana Bhashani Science and Technology University
Santosh-1902, Tangail, Bangladesh
Email: amirulce10022@gmail.com, m.hasan2006@yahoo.com, zia@iut-dhaka.edu, mm73rahman@gmail.com

*Abstract*—Recently frequent and sequential pattern mining algorithms have been widely used in the field of software engineering to mine various source code or specification patterns. In practice software evolves from one version to another is needed for providing extra facilities to user. This kind of task is challenging in this domain since the database is usually updated in all kinds of manners such as insertion, various modifications as well as removal of sequences. If database is optimized then this optimized information will help developer in their development process and save their valuable time as well as development expenses. Some existing algorithms which are used to optimize database but it does not work faster when database is incrementally updated. To overcome this challenges an efficient algorithm is recently introduce, called the Canonical Order Tree that captures the content of the transactions of the database and orders. In this paper we have proposed a technique based on the Canonical Order Tree that can find out frequent patterns from the incremental database with speedy and efficient way. Thus the database will be optimized as well as it gives useful information to recommend software developer.

*Keywords—Frequent pattern mining, Canonical Order tree, Incremental mining, AFPIM (Adjust Frequent Pattern Incremental mining).*

## I. INTRODUCTION

The problem of mining association rules and the more general problem of searching frequent patterns from huge databases have been the subject of numerous studies. These studies can be classified into two categories. First category is the functionality where the major question is what kind of rules or patterns are needed to compute. While some studies [1], [2], [3] in this category are considered as data mining exercise in isolation, whereas the data mining can work best with the database management system [4], [5] and the thing that human use. Second category is performance where the major question is considered that how to determine the association rules or frequent patterns as perfectly as possible. Studies in this category again classified into several subgroups. This subject of mining frequent pattern also used in software development or update process. Software development process for a system is originally done by writing a program based on the specific design specification. At present it can be done by various techniques. The main purpose of the software development or update is to speed up the development process. For this developers in the software industries usually create new methods, attributes or modifies existing or reuse previously define methods, attributes. One of the processes to develop software is reusing the existing libraries and frameworks. Developers can be highly benefited if they get appropriate guidelines and accurate recommendations. It can decrease development cost as well as save valuable time. Scanning and Recommending Application Programming Interface (API) is a process to search relevant items from an existing repository.

This process is used widely across all over the world. Some approaches of these API searching processes are based on searching online. Developers generally search for API usage pattern using online repositories provided by various tools.

Code search engines often deals with the extra large files. The extra large files often contain a large number of API methods, attributes. Developers face the difficulty is to choose the appropriate API methods, attributes to use. As the database contains a large amount of data and takes a long time to find appropriate methods, attributes and also it does not provide specific frequent pattern of attributes. However, to solve the problem of finding frequent pattern an efficient algorithm called Canonical Order tree [6], [7] can be applied and it perform well than some existing algorithms. This task of finding frequent pattern can be applied in other field like optimization of database to gain some valuable information and based on this information provide appropriate guidelines to developers for evolving software from one version to another. It is needed for software developer to take decision which portion of existing version is needed or which portion can be ignored and modified to make the software more user friendly. Such information can be obtained from software database and normally that portion which is frequent is needed to add with the update version of the software. Based on this information database of the software is optimized. But if there are any system or technique which helps to gain this optimized information of the software database that will be very useful for software developer. In this paper we show this technique and the experimental result shows that our proposed technique are performed better than the existing algorithm [8], [9] based on the criteria of time complexity.

## II. BACKGROUND AND RELATED WORKS

There are some existing algorithms which mine frequent pattern and optimize database as perfectly as possible. Studies in this category again can be classified into several sub- groups. The first subgroup is made of fast algorithms based on the level wise Apriori framework [10] and second sub- group focuses on the performance enhancement mechanism like hashing and segmentation [11], [12] for speeding up Apriori based algorithms. In summary, those Apriori based incremental mining algorithms can not be easily adoptable to FP-tree based [13], [14] incremental mining. To overcome the situation the Frequent/Large patterns mining (FELINE) with the Compressed and Arranged Transactional Sequences (CATS) tree [8] was invented. But it was mainly designed for interactive mining, where the build once, mine many principle holds. However, such mechanism does not necessarily hold for incremental mining. The another algorithm called AFPIM algorithm [9] was proposed to reduce this problem but it is unable to eliminate the task of rescanning of the database when it is change dynamically. In our proposed technique which is based on canonical order tree can mine frequent pattern from the incremental database and rescanning is not needed when database is changed dynamically.

## III. PROPOSED TECHNIQUE

We use the sample database which is taken from open source repository and apply our proposed technique to find the frequent pattern and thus optimize the database of the software. Here we have considered a sample database called google Maps API in version 2 and version 3. The following table and figure shows our implementation of the techniques. Here it can be seen that the frequent pattern (e.g. methods, properties, events use in various classes) of the database of the software in version 2 is also used in updated version 3. Here the minimum support we use 50% that means minimum support value=2. In V2 frequent pattern are mouseout, mouseover, clickable and getBounds(). In V3 the frequent pattern are clickable, getBounds(), mouseover, getmap(), rightclick and visible.

### A. Find frequent items

Here the term frequent item means the item which satisfying the minimum support.

### B. Applying canonical order tree to find frequent item

We have applied canonical order tree as sequential or frequent pattern mining algorithm to optimize the item of database Table I and III. The result is represented with Figure 1 and 2 respectively. It will scan the items as per the minimum support and sort out the expected items.

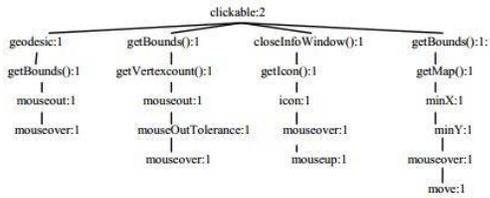

Fig. 1. Implementation of Canonical Order Tree for Google Maps API V2

TABLE I.    SAMPLE DATABASE OF GOOGLE API V2

| Serial | Transaction | Item set |
|---|---|---|
| Item1 | google.maps.Map | getMap(),getBounds(),mouseover,minX,minY,move |
| Item2 | google.maps.Marker | closeInfoWindow(),getIcon(),icon,title,mouseover,mouseup |
| Item3 | google.maps.Polyline | getBounds(),mouseout,mouseover,clic kable |
| Item4 | google.maps.Polygon | getBounds(),mouseout,mouseover,clic kable |

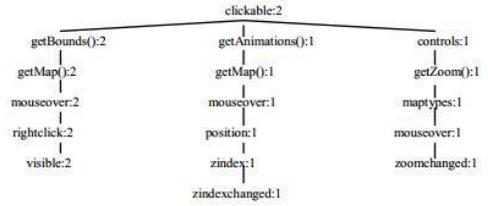

Fig. 2. Implementation of Canonical Order Tree for Google Maps API V3

TABLE II.    OPTIMIZED SAMPLE DATABASE OF GOOGLE MAP API V2

| Serial | Transaction | Item set |
|---|---|---|
| Item1 | google.maps.Map | getBounds(),mouseover |
| Item2 | google.maps.Marker | Mouseover |
| Item3 | google.maps.Polyline | getBounds(),mouseout,mouseover,clic kable |
| Item4 | google.maps.Polygon | getBounds(),mouseout,mouseover,clic kable |

TABLE III.    SAMPLE DATABASE OF GOOGLE MAP API V3

| Serial | Transaction | Item set |
|---|---|---|
| Item1 | google.maps.Mapm | getZoom(),setOptions(),mapTypes,zoomchanged,mouseover,controls |
| Item2 | google.maps.Marker | getAnimation(),getMap(),position,zIndex,mouseover,zindexchanged |
| Item3 | google.maps.Polyline | getMap(),getBounds(),visible,rightclick,mouseover,clickable |
| Item4 | google.maps.Polygon | getMap(),getBounds(),visible,rightclick,mouseover,clickable |

TABLE IV.    OPTIMIZED SAMPLE DATABASE OF GOOGLE MAP API V3

| Serial | Transaction | Item set |
|---|---|---|
| Item1 | google.maps.Map | mouseover |
| Item2 | google.maps.Marker | getMap(),mouseover |
| Item3 | google.maps.Polyline | getMap(),getBounds(),visible,rightclick,mouseover,clickable |
| Item4 | google.maps.Polygon | getMap(),getBounds(),visible,rightclick,mouseover,clickable |

The algorithm is applied on the Table I and after optimization Table II is generated.

The algorithm is applied on the Table III and after optimization Table IV is generated.

## IV. ALGORITHM ANALYSIS

In our proposed approach Canonical Order Tree [6], [7] is applied over the other Frequent Mining Algorithms. From the investigation it is found that Canonical Order Tree has the maximum efficiency comparing with other algorithms like GSPAN, PrefixSpan, SPADE etc. The implementation of Canonical order tree is given here.

Step-1: Input transaction number, data item sets in 2D array, minimum support.

Here the transaction number represent items of the database table. Minimum support is an integer value and depends on this value the canonical order tree algorithm find the sequential or frequent item from the database. Data sets input in two dimensional array. Here two dimension is represent database table row and column. Step 1 is illustrated in the Figure 3.

```
trasaction_number = input.nextInt();
t=t+trasaction_number;
for(k=n+1;k<=t;k++)
{m= input.nextInt();
 for (i=1;i<=m;i++)
 a[i]=inData.readLine();
 for (i=1;i<=m;i++)
 for (j=i+1;j<=m;j++)
 { q= a[i].compareTo(a[j]);
   if(q>0) {
   temp = new String(a[i].toString());
   a[i] = new String(a[j].toString());
   a[j] = new String(temp.toString());
 }}
```

Fig. 3. Code Snippet of Primary Input

Step-2: Sorted data items into canonical order (for example lexicographic order). Step 2 is illustrated in the Figure 4.

```
for(i=1;i<=p1;i++)
{ for(j=1;j<=p1;j++)
  {if(i==j)continue;else
  {k=1;while(k<=l)
  {q = b[i].equalsIgnoreCase(c[k]);
    k++;}
```

Fig. 4. Code Snippet of Canonical Orde

Step -3: Support count of every items of Sorted data item. Step 3 is illustrated in the Figure 5.

```
if(flag==1)
{q= b[i].equalsIgnoreCase(b[j]);
 if(q==true){count++;}}
```

Fig. 5. Code Snippet of the support count of sorted data items

Step- 4: Checked if item set satisfied minimum support.

If yes then item set is stored in the array and considered it as frequent item. If no then ignore that item set. Step 4 is illustrated in the Figure 6.

```
if(flag==1)
{q= b[i].equalsIgnoreCase(b[j]);
 if(q==true){count++;
 c[l] = new String(b[i].toString());
 l++;}}
```

Fig. 6. Code Snippet of checking item satisfy minimum support

Step 5: Checked if all sorted item sets is checked?

If yes then frequent item set is stored in the array and ready to print.

If no then continue from step 3 to 5.

## V. EVALUATION

To evaluate the applicability and the efficiency of our proposed approach, we have compared it with the other existing algorithms like AFPIM, FELINE. Before, analyzing the result a dedicated experimental environment is set up.

### A. Environmental Set up

The result found are simulated from WEKA [15] MATLAB, Netbeans IDE and Turbo C compiler in system of core-i5

```
if(count>=min_support)
{System.out.println("\n");
 System.out.print(" "+b[i]);}
```

Fig. 7. Code Snippet of checking all items of database

processor with 4GB RAM along with the internet connectivity of local operator.

### B. Time Complexity Evaluation

The algorithm is implemented in Turbo C compiler as well as Netbeans7.3 compiler(first 3 is Netbean and last 3 is C compiler and time calculate in millisecond).

TABLE V. TIME COMPLEXITY EVALUATION

| Task | Canonical order tree | AFPIM | FELINE |
|---|---|---|---|
| 1 | 255060 | 373932 | 405429 |
| 2 | 299515 | 373998 | 773455 |
| 3 | 373752 | 573997 | 999465 |
| 4 | 255165 | 405419 | 573999 |
| 5 | 299717 | 433799 | 905997 |
| 6 | 373852 | 583999 | 999715 |

Here the Y-axis represent time in ms and X-axis represent iteration of applying three techniques.

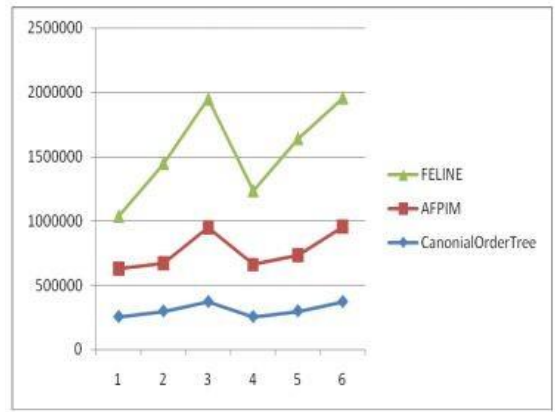

Fig. 8. Runtime evaluation in graphical representation

From the above runtime evaluation Figure it can be seen that our propose technique based on canonical order tree is comparatively more than the others.

### C. Empirical Evaluation

To see that if the proposed approach is helpful to software developer, an experimental investigation is held through empirical evaluation. Before starting up the empirical evaluation, four different teams consisting of three members are formed, and they are said to work with the different systems environment including the Can Tree Approach along with the other existing algorithm.

The result is given below.

Here Y-axis represent error and X-axis represent team number

From the above Figure it can be seen that the proposed approach brings less error comparatively than others.

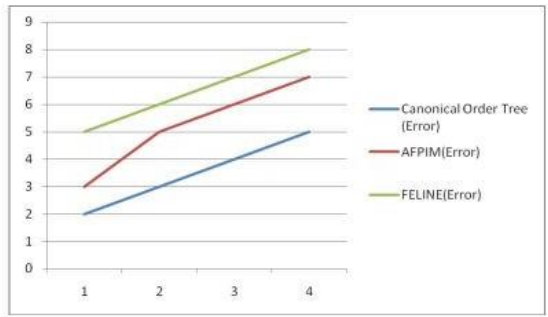

Fig. 9. Graphical representation of empirical evaluation

TABLE VI. EMPIRICAL EVALUATION

| Team | Canonical order tree (Error) | AFPIM(Error) | FELINE(Error) |
|---|---|---|---|
| 1 | 2 | 3 | 5 |
| 2 | 3 | 5 | 6 |
| 3 | 4 | 6 | 7 |
| 4 | 5 | 7 | 8 |

## VI. THREAT TO THE EVALUATION AND LIMITATION

Here in the experimental evaluation google API is considered that is not standard repository. If different repository is used the result may be varied. If team member of the empirical evaluation is changed the result may be changed. Learning curve of the particular team member influences the evaluation result, as seen from the experimental evaluation.

## VII. CONCLUSION & FUTURE WORK

Software development is getting changed so rapidly. Existing repository can help software developers. If the existing repository is optimized then developers can be guided accurately. The frequent pattern, found from the incremental database using the proposed approach is more relevant and convenient than the others. Based on this frequent pattern database is optimized and this optimized information recommend software developer. We still have been working on the proposed approach to extend its applicability for the professional uses, through real time implementation, the proposed approach also can be applied along with the application of graph mining algorithm. The repository we use is not stan- dard so we want to work with standard repository. Also we want to work with better simulation environment than current simulation environment.